\documentclass[]{aastex631}

\newcommand{\be}{\begin{equation}}
\newcommand{\ee}{\end{equation}}

\begin{document}

\title[Galaxy Zoo UKIDSS]{Galaxy Zoo: Morphologies based on UKIDSS NIR Imaging for 71,052 Galaxies}

\author[0000-0003-0846-9578]{Karen L. Masters}
\affiliation{Departments of Astronomy and Physics, Haverford College, 370 Lancaster Ave., Haverford, PA 19041, USA}
\email{klmasters@haverford.edu}

\author{Melanie Galloway}
\altaffiliation{Now at ZEISS Industrial Metrology}
\affiliation{School of Physics and Astronomy, University of Minnesota, 116 Church Street SE, Minneapolis, MN 55455, US}

\author[0000-0002-1067-8558]{Lucy Fortson}
\affiliation{School of Physics and Astronomy, University of Minnesota, 116 Church Street SE, Minneapolis, MN 55455, US}

\author{Chris J. Lintott}
\affiliation{Department of Physics, University of Oxford, Denys Wilkinson Building, Keble Road, Oxford, OX1 3RH, UK}

\author{Mike Read}
\affiliation{Royal Observatory Edinburgh, Blackford Hill,  Edinburgh, EH9 3HJ, UK}

\author[0000-0002-9136-8876]{Claudia Scarlata}
\affiliation{School of Physics and Astronomy, University of Minnesota, 116 Church Street SE, Minneapolis, MN 55455, US}

\author[0000-0001-5882-3323]{Brooke Simmons}
\affiliation{Department of Physics, Lancaster University, Lancaster LA1 4YB, UK}

\author[0000-0002-6408-4181]{Mike Walmsley}
\affiliation{Dunlap Institute for Astronomy and Astrophysics, University of Toronto, 50 St. George Street, Toronto, ON M5S 3H4, Canada}

\author[0000-0002-3654-3504]{Kyle Willett}
\altaffiliation{Now at Amazon}
\affiliation{School of Physics and Astronomy, University of Minnesota, 116 Church Street SE, Minneapolis, MN 55455, US}

\begin{abstract}
We present morphological classifications based on Galaxy Zoo analysis of 71,052 galaxies with imaging from the United Kingdom Infrared Telescope Infrared Deep Sky Survey (UKIDSS). Galaxies were selected out of the Galaxy Zoo 2 (GZ2) sample, so also have $gri$ imaging from the Sloan Digital Sky Survey. An identical classification tree, and vote weighting/aggregation was applied to both UKIDSS and GZ2 classifications enabling direct comparisons. With this Research Note we provide a public release of the GZ:UKIDSS morphologies and discuss some initial comparisons with GZ2.
\end{abstract}

\keywords{} 

\section{Introduction} \label{sec:intro}

Historically, most morphological classification of galaxies has been conducted using optical images. In the photographic era, B-band images were the primary source \citep[e.g.][]{Hubble1926,Sandage1961,DV1963}. More recent morphological catalogs also derive their classifications from rest-frame optical images, e.g. B-band \citep{DV1991}, ACS I-F814W  \citep{scar}, Sloan Digital Sky Survey (SDSS) $g$-band \citep{Fukugita,Nair} or SDSS $gri$ color-composite \citep{Lintott,Willett}. 

Optical galaxy flux is dominated by young, hot stars, highlighting structures where star formation is ongoing. Optical light is also impacted by extinction due to dust, which can obscure features. Longer wavelengths are less impacted by these effects. Near infra-red (NIR) galaxy flux is dominated by lower mass, redder stars, thus should reveal the underlying “stellar backbone” of galaxies. Indeed, studies of individual or small numbers of galaxies have have found significant differences between optical and NIR morphology
\citep{Hack,Thronson,Block91,Block1994,BlockandP1999}; although, in larger samples IR morphology has been found to be well-correlated with optical morphology \citep{Esk2002,Buta2010}. 

With this Research Note, we publish Galaxy Zoo classifications for a sample of 71,052 galaxies with imaging from the United Kingdom Infrared Telescope (UKIRT) Infrared Deep Sky Survey (UKIDSS; \citealt{Lawrence2007}). These data have been used to investigate the difference between optical \citep[Galaxy Zoo 2 or GZ2; ][]{Willett} and NIR morphology \citep{Galloway,Warrick}. 

\section{UKIDSS Imaging and Sample Selection}

UKIDSS \citep[][]{Lawrence2007,Warren2007}  was a large imaging program on UKIRT from 2005-2012. The Large Area Survey (LAS) of UKIDSS, obtained ZYJHK imaging across almost 4000 sq degs at high Galactic latitude.  The UKIDSS Galaxy Zoo sample is comprised of 71,052 galaxies with LAS imaging selected out of GZ2 (previously classified based on SDSS $gri$ images, \citealt[][]{Willett}). GZ2 selected the brightest 25\% of galaxies (magnitude $m_r<17$ and angular size {\tt petroR90\_r}$>3$\arcsec) from the SDSS Main Galaxy Sample \citep{Strauss2002}. 

 UKIDSS image cutouts were generated using YJK data (1--2.4$\mu$m) implementing the \citet{Lupton2004} algorithm to create color images with a comparable color balance to SDSS $gri$ (0.48--0.76$\mu$m).\footnote{The same algorithm is implemented in the {\tt ColourImage} service at {\tt http://wsa.roe.ac.uk/}} The resolution of both sets of images are comparable (1.2-1.4\arcsec ~with a pixel scale around 0.4\arcsec), however SDSS imaging is typically 1 mag deeper than UKIDSS (SDSS has typical magnitude limits of 22.2-21.3 across $gri$ compared to 20.3-21.1 across YJK in UKIDSS; with bluest bands being deepest in both surveys). Example images from each survey are shown in Figure \ref{figure}a,b.

\section{Galaxy Zoo UKIDSS Classifications} 

Morphological classifications for this UKIDSS sample were collected via Galaxy Zoo\footnote{www.galaxyzoo.org} from October 2013-May 2014. More than 80,000 individual volunteers contributed classifications. The classification tree and methods to count and weight raw votes by user consistency were both identical to that in GZ2 \citep{Willett}, to allow for a direct comparison. The reader is referred to \citet{Willett} for more details. 

The UKIDSS Galaxy Zoo classification catalog is available at \href{https://data.galaxyzoo.org/}{\tt data.galaxyzoo.org} and both images and catalog are available at HuggingFace\footnote{\url{https://huggingface.co/datasets/mwalmsley/gz_ukidss}}, as part of a large Galaxy Zoo multi-campaign dataset used to train Zoobot \citep{Walmsley2023}.

\section{NIR Compared to Optical Morphologies}

\begin{figure}
\begin{center}
\includegraphics[scale=.65]{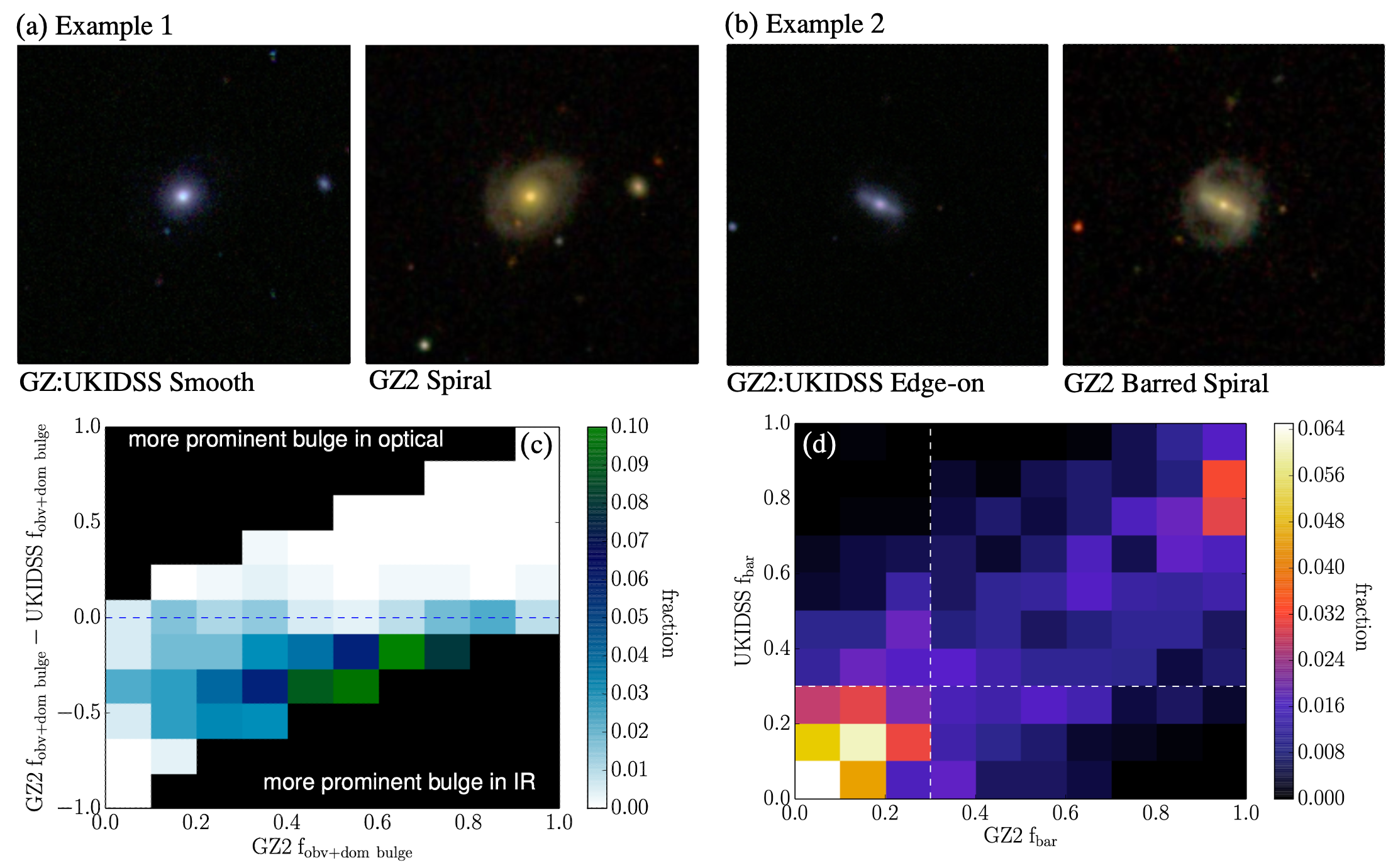}
\caption{Panels (a,b) show two sets of example images. Panel (c) is a figure from \citet{Galloway} showing how in 502 spiral galaxies $f_{\rm obv+dom}$ increases between GZ2 and GZ:UKIDSS and panel (d) is also from \citet{Galloway}, showing the correlation between GZ2 and UKIDSS bar vote fractions for 1,107 galaxies.  \label{figure}}
\end{center}
\end{figure}

\citet{Galloway} made a comparison of morphologies of 6,484 bright nearby galaxies clearly visible in both GZ2 (SDSS) and GZ:UKIDSS. This selection was an attempt to mitigate the impact of shallower UKIDSS imaging. \citet{Galloway} found that galaxies as seen in UKIDSS tended to have lower vote fractions for visible features ($f_{\rm features}$) and larger vote fractions for obvious or dominant bulges ($f_{\rm obv+dom}$, see Figure \ref{figure}c). Thus galaxies in UKIDSS had slightly earlier-appearing morphologies, consistent with the NIR imaging revealing more of the older stellar populations than optical. Where spirals were identified in both sets of images, they tended to appear smoother and looser in UKIDSS, but often, spirals visible in SDSS were invisible in the UKIDSS imaging. Bars ($f_{\rm bar}>0.3$) were not significantly more obvious in the UKIDSS imaging than SDSS (see Figure \ref{figure}d).

\citet{Galloway} concluded that the majority of the differences in morphology between UKIDSS and SDSS imaging were driven by the difference in depth of the images compared to the typical brightness of galaxies at those wavelengths (as illustrated in Figure~\ref{figure}a,b). For a more useful comparison of NIR to optical morphologies in a large sample, analysis will be needed on a large sample with much deeper NIR imaging. This should be possible in the near future with the Galaxy Zoo: Euclid project \citep{EuclidCollaboration2024}, which has both optical and YJH imaging.

\begin{acknowledgments}
This publication uses data generated via the Zooniverse.org platform, development of which is funded by generous support, including a Global Impact Award from Google, and by a grant from the Alfred P. Sloan Foundation. This publication has been made possible by the participation of more than 80,000 volunteers in the Galaxy Zoo project. 

We thank Steve Warren (Imperial) for support and assistance with accessing UKIDSS imaging and explanations of common artifacts. 

\end{acknowledgments}

\vspace{5mm}
\facilities{UKIDSS, SDSS}


\end{document}